\documentclass[referee]{aa}
\begin{document}

\title{
Maximum mass and radius of strange stars in the linear approximation of the EOS
} 
\author{T. Harko and K. S. Cheng}   
\institute{
Department of Physics, The University of Hong Kong, Pok Fu Lam Road, Hong Kong, P. R. China
}
\date{Received 9 August 2001/Accepted 4 February 2002}
\authorrunning{Harko \& Cheng}
\titlerunning{Maximum Mass and Radius for Linear EOS}

\maketitle

\begin{abstract}
The Chandrasekhar limit for strange stars described by a linear
equation of state (describing quark matter with density-dependent quark masses) is evaluated.
The maximum mass and radius of the star depend on the fundamental constants
and on the energy density of the quark matter at zero pressure. By comparing the expression for
the mass of the star with the limiting mass formula for a relativistic degenerate
stellar configuration one can obtain an estimate of the mass of the strange quark.


\end{abstract}
 
\section{Introduction}

One of the most important characteristics of compact relativistic
astrophysical objects is their maximum allowed mass. The maximum mass is
crucial for distinguishing between neutron stars and black holes in compact
binaries and in determining the outcome of many astrophysical processes,
including supernova collapse and the merger of binary neutron stars. The
theoretical value of the maximum mass for white dwarfs and neutron stars was
found by Chandrasekhar and Landau and is given by $M_{\max }\sim \left( 
\frac{\hbar c}{G}m_{B}^{-4/3}\right) ^{3/2}$ (\cite{ShTe83}), where $m_{B}$ is
the mass of the baryons (in the case of white dwarfs, even pressure comes from
electrons; most of the mass is in baryons). Thus, with the exception of
composition-dependent numerical factors, the maximum mass of a degenerate
star depends only on fundamental physical constants. The radius $R_{\max }$
of the degenerate star obeys the condition $R_{\max }\leq \frac{\hbar }{mc}%
\left( \frac{\hbar c}{Gm_{B}^{2}}\right) ^{1/2}$, with $m$ being the mass of either
electron (white dwarfs) or neutron (neutron stars) (\cite{ShTe83}). White dwarfs are supported
against of gravitational collapse by the degeneracy pressure of electrons
whereas for neutron stars this pressure comes mainly from the nuclear force between nucleons (\cite{ShTe83}). 
For non-rotating neutron stars with the central pressure at their center tending to the
limiting value $\rho _{c}c^{2}$, an upper bound of around $3M_{\odot}$  has been
found (\cite{Rh74}).

The quark structure of the nucleons, suggested by quantum cromodynamics,
opens the possibility of a hadron-quark phase transition at high densities
and/or temperatures, as suggested by Witten (\cite{Wi84}). In the theories of
strong interaction, quark bag models suppose that breaking of physical vacuum
takes place inside hadrons. As a result, vacuum energy densities inside and
outside a hadron become essentially different and the vacuum pressure on the
bag wall equilibrates the pressure of quarks, thus stabilizing the system. If
the hypothesis of the quark matter is true, then some neutron stars
could actually be strange stars, built entirely of strange matter (\cite{Al86}%
, \cite{Ha86}). For a review of strange star properties, see (\cite{Ch98}).

Most of the investigations of quark star properties have been done within
the framework of the so-called MIT bag model. Assuming that interactions of
quarks and gluons are sufficiently small, neglecting quark masses and
supposing that quarks are confined to the bag volume (in the case of a bare
strange star, the boundary of the bag coincides with the stellar surface), the
energy density $\rho c^{2}$ and pressure $p$ of a quark-gluon plasma at
temperature $T$ and chemical potential $\mu _{f}$ (the subscript $f$
denotes the various quark flavors $u$, $d$, $s$ etc.) are related, in the MIT
bag model, by the equation of state (EOS) (\cite{Ch98}) 
\begin{equation}\label{1}
p=\frac{(\rho -4B)c^{2}}{3},  
\end{equation}
where $B$ is the difference between the energy density of the perturbative
and non-perturbative QCD vacuum (the bag constant). Equation (\ref{1}) is
essentially the equation of state of a gas of massless particles with
corrections due to the QCD trace anomaly and perturbative interactions.
These corrections are always negative, reducing the energy density at given temperature
by about a factor of two (\cite{Fa84}). For quark stars
obeying the bag model equation of state (\ref{1}) the Chandrasekhar limit
has been evaluated, from simple energy balance relations, in (\cite{Ba00}). In
addition to fundamental constants, the maximum mass also depends on the bag
constant.    

More sophisticated investigations of quark-gluon interactions have shown
that Eq. (\ref{1}) represents a limiting case of more general equations of
state. For example, MIT bag models with massive strange quarks and lowest-
order QCD interactions lead to some corrections terms in the equation of
state of quark matter. Models incorporating restoration of chiral quark
masses at high densities and giving absolutely stable strange matter can no
longer be accurately described by using Eq. (\ref{1}). On the other hand, in
models in which quark interaction is described by an interquark potential
originating from gluon exchange and by a density-dependent scalar potential
which restores the chiral symmetry at high densities (\cite{De98}), the equation of state $P=P\left( \rho \right) $ can be well
approximated by a linear function in the energy density $\rho $ (\cite{Go00}).
It is interesting to note that Frieman
and Olinto (\cite{Fr89}) and Haensel and Zdunik (\cite{Ha89})
have already mentioned the approximation of the EOS by a linear function (see also
Prakash et al. (\cite{Pr90}), Lattimer et al (\cite{La90}). Recently
Zdunik (\cite{Zd00}) has studied the linear approximation of the
equation of state, obtaining all parameters of the EOS as polynomial
functions of strange quark mass, the QCD coupling constant and bag constant. The
scaling relations have been applied to the determination of the maximum
frequency of a particle in a stable circular orbit around strange stars.

It is the purpose of this paper to obtain, by using a simple phenomenological approach
(which is thermodynamical in its essence), the maximum mass and
radius (the Chandrasekhar limits) for strange stars obeying a linear equation of
state. Of course the maximum mass of compact astrophysical objects is a
consequence of General Relativity and not of the character of motion of matter
constituents. However, the formulae for maximum mass and radius, due to their simple analytical form,
give a better insight into the underlying physics of quark stars, also allowing us
to obtain some results which cannot be obtained by numerical methods.
For example, from the obtained relations one can find the scaling relations for the maximum
mass and radius of strange stars in a natural way.

The present paper is organized as follows. The maximum mass and radius of quark stars with a general
linear equation of state is obtained in Section 2. In Section 3 we discuss our results
conclude the paper.

\section{Maximum mass and radius for strange  stars in the linear
approximation of the EOS}

We assume that the strange star obeys an equation of state
that can be obtained  by interpolation with a linear function of density in
the form: 
\begin{equation}\label{2}
p=a\left( \rho -\rho _{0}\right) c^{2},  
\end{equation}
where $a$ and $\rho _{0}$ are non-negative constants. $\rho _{0}$ is the
energy density at zero pressure.

Such an equation of state has been proposed
mainly to describe the strange matter built of $u$, $d$ and $s$ quarks (\cite{Go00}, \cite{Zd00}).
The physical consistency of the model requires $\rho _{0}>0$.

The particle number density and the chemical potential corresponding to EOS (\ref{2}) are given
respectively by (\cite{Zd00})
\begin{equation}\label{3}
n\left( p\right) =n_{0}\left( 1+\frac{a+1}{a}\frac{p}{\rho _{0}c^{2}}\right)
^{1/\left( a+1\right) },  
\end{equation}
\begin{equation}\label{4}
\mu \left( p\right) =\mu _{0}\left( 1+\frac{a+1}{a}\frac{p}{\rho _{0}c^{2}}%
\right) ^{a/\left( a+1\right) },  
\end{equation}
where $n_{0}$ is the particle number density at zero pressure and $\mu
_{0}=\rho _{0}c^{2}/n_{0}$.

The parameters $a$ and $\rho _{0}$ can be calculated, for realistic equations
of state, by using a least squares fit method (\cite{Go00}, \cite{Zd00}).
For the equations of state incorporating restoration of chiral quark masses at high densities proposed
in Dey et al. (\cite{De98}) one obtains the values $a=0.463$, $\rho _{0}=1.15\times 10^{15}g/cm^3$ and
$a=0.455$, $\rho _{0}=1.33\times 10^{15}g/cm^3$, respectively (\cite{Go00}). 
The standard bag model corresponds to $a=0.333$ and $\rho _{0}=4\times 10^{14}g/cm^3$ (\cite{Ch98}).

From Eqs. (\ref{3})-(\ref{4}) it follows that the particle number and
chemical potential are related by the equation
\begin{equation}\label{5}
\mu =\frac{\rho _{0}c^{2}}{n_{0}^{a+1}}n^{a}.  
\end{equation}

From the numerical studies of strange star models we know
that the density profile of this type of astrophysical object 
is quite uniform (\cite{Gl96}). Therefore we can approximate $n\approx N/V$, which leads to
\begin{equation}\label{6}
\frac{\mu }{\mu _{0}}=\left( \frac{4\pi }{3}\right) ^{-a}\left( \frac{N}{%
n_{0}}\right) ^{a}R^{-3a},  
\end{equation}
where $N$ is the total number of particles in a star of radius $R$ and
volume $V$.

With the use of Eqs. (\ref{2})-(\ref{6}) one obtains the energy density of
the star in the form
\begin{equation}\label{7}
\rho =\frac{\rho _{0}}{a+1}\left( \frac{4\pi }{3}\right) ^{-a-1}\left( \frac{%
N}{n_{0}}\right) ^{a+1}R^{-3\left( a+1\right) }+\frac{a}{a+1}\rho _{0}.
\end{equation}

The total mass $M$ of the star is defined according to $M=4\pi
\int_{0}^{R}\rho r^{2}dr\approx \frac{4\pi }{3}\rho R^{3}$ and is given by
\begin{equation}\label{8}
M=\frac{\rho _{0}}{a+1}\left( \frac{4\pi }{3}\right) ^{-a}\left( \frac{N}{%
n_{0}}\right) ^{a+1}R^{-3a}+\frac{4\pi }{3}\frac{a}{a+1}\rho _{0}R^{3},
\end{equation}
where we assumed that the energy density is approximately constant inside the star.

Extremizing the mass with respect to the radius $R$ by means of  $\partial
M/\partial R=0$ gives the relation
\begin{equation}\label{9}
\frac{\rho _{0}}{a+1}\left( \frac{4\pi }{3}\right) ^{-a}\left( \frac{N}{n_{0}%
}\right) ^{a+1}R^{-3a}=\frac{4\pi }{3}\frac{1}{a+1}\rho _{0}R^{3}.  
\end{equation}

Substituting Eq. (\ref{9}) into Eq. (\ref{8}) we obtain the maximum mass of
the strange star in the linear approximation of the EOS:
\begin{equation}\label{10}
M=\frac{4\pi }{3}\rho _{0}R^{3}.  
\end{equation}

This expression is very similar to the expression for the maximum mass of
the quark star obtained assuming that the star is composed of three-flavour
masslesss quarks, confined in a large bag (\cite{Ba00}, \cite{ChHa00}). From a physical point of view,
Eq. (\ref{10}) describes a uniform density zero pressure stellar type configuration.

The maximum equilibrium radius corresponds to a minimum total
energy of the star (including the gravitational one), for any radius. For
ordinary compact stars, the mass is entirely due to baryons, and the
corresponding (Newtonian) gravitational potential energy is of the order $%
E_{G}\sim -\alpha GM^{2}/R$ ($\alpha =-3/5$ for constant density Newtonian
stars). For quark stars, assumed to be formed of massless quarks, the total
mass can be calculated from the total (thermodynamic as well as confinment)
energy in the star. One possibility for the estimation of the gravitational
energy per fermion is to define an effective quark mass incorporating all
the energy contributions (\cite{Ba00}).

The gravitational energy per particle (the strange star is assumed to be
formed from fermions) is
\begin{equation}
E_{G}=-\frac{GMm_{eff}}{R},
\end{equation}
where $m_{eff}$ is the effective mass of the particles inside the star,
incorporating also effects such as quark confinment. For a star with $N$
particles one can write $M=Nm_{eff}=\rho _{0}V$, or $m_{eff}=\rho _{0}/n$.
On the other hand one can assume $\mu =\rho _{0}/2n$ (\cite{Ba00}), leading
to $m_{eff}=2\mu /c^{2}$. Hence, with the use of Eqs. (\ref{6}), (\ref{9})
and (\ref{10}) we can express the gravitational energy per particle as
\begin{equation}
E_{G}=-2\left( \frac{4\pi }{3}\right) ^{2}G\frac{\rho _{0}^{2}}{N}R^{5}.
\end{equation}

The energy density per particle of the fermions follows from Eq. (\ref{7})
and is given by:
\begin{equation}
E_{F}=\frac{4\pi }{3}\frac{1}{a+1}\frac{\rho _{0}}{N}R^{3}.
\end{equation}

The total energy $E$ per particle is
\begin{equation}
E=\frac{4\pi }{3}\frac{1}{a+1}\frac{\rho _{0}}{N}R^{3}-2\left( \frac{4\pi }{3%
}\right) ^{2}G\frac{\rho _{0}^{2}}{N}R^{5}.
\end{equation}

Extremizing the total energy with respect to the radius (with the total
particle number kept constant), $\left( \frac{\partial E}{\partial R}\right)
_{N=const.}=0$, it follows that the maximum radius of the equilibrium
configuration in the linear approximation of the EOS is given by:
\begin{equation}\label{12}
R_{max}=R_{0}\frac{c}{\sqrt{\pi \left( a+1\right) G\rho _{0}}}.  
\end{equation}

The maximum mass of the star can be calculated from Eq. (\ref{10}) and is: 
\begin{equation}\label{13}
M_{max}=\frac{4}{3}\frac{R_{0}^{3}}{\left( a+1\right) ^{3/2}}\frac{c^{3}}{G}%
\frac{1}{\sqrt{\pi G\rho _{0}}}.
\end{equation}

In Eqs. (\ref{12}) and (\ref{13}) $R_{0}$ is a numerical factor of the order 
$R_{0}\approx 0.474$.

The maximum radius of the quark star given by Eq. (\ref{12}) is the radius corresponding
to the maximum mass. On the other hand for the existing models of strange stars,
the configuration with maximum mass has a radius which is lower than the maximum radius.
For example, for strange stars described by the bag model equation of state, the maximum
radius is $11.40 km$, while the radius corresponding to the maximum mass is $10.93 km$, which is
$4\%$ lower than the maximum radius. This difference is neglected in Eq. (\ref{12}).

The maximum mass and radius of the star are strongly
dependent on the numerical value of the coefficient $R_{0}$ and estimations based on other
physical models could lead to different numerical estimates of the limiting values of the basic
parameters of the static strange stars.

\section{Discussions and final remarks}

In the present paper we have shown that there is a maximum mass and radius (the
Chandrasekhar limits) for quark stars whose
equation of state can be approximated by a linear function of
the density. We have also obtained the explicit expressions for $M_{max}$ and $R_{max}$.

With respect to the scaling of
the parameter $\rho _{0}$ of the form $\rho _{0}\rightarrow k\rho _{0}$, the
maximum mass and radius have the following scaling behaviors:
\begin{equation}\label{14}
R_{max}\rightarrow k^{-1/2}R_{\max }, M_{\max }\rightarrow k^{-1/2}M_{\max }.
\end{equation}

For the maximum mass of the strange stars this scaling relation has also been
found from the numerical study of the structure equations
in the framework of the bag model (\cite{Wi84}, \cite{Ha86}).

A rescaling of the parameter $a$ of the form $a+1\rightarrow K\left(
a+1\right) $, with $\rho _{0}$ unscaled, leads to a transformation of the
radius and mass of the form
\begin{equation}
R_{max }\rightarrow K^{-1/2}R_{max }, M_{max }\rightarrow K^{-3/2}M_{max
}.
\end{equation}

A simultaneous rescaling of both $a$ and $\rho _{0}$, with $a+1\rightarrow
K\left( a+1\right) ,\rho _{0}\rightarrow k\rho _{0}$ gives
\begin{equation}
R_{max }\rightarrow k^{-1/2}K^{-1/2}R_{max }, M_{max }\rightarrow
k^{-1/2}K^{-3/2}M_{max }.
\end{equation}

The maximum mass and radius of strange stars with linear EOS is strongly dependent
on the numerical value of $\rho _{0}$, the mass decreasing with increasing $\rho _{0}$. For $\rho _{0}=4B$,
with the bag constant $B=10^{14}g/cm^3$ ($56MeVfm^{-3}$) we obtain                                             
$M_{max}=1.83M_{\odot}$, a value that must be compared to the value $M_{max}=2M_{\odot}$
obtained by numerical methods (\cite{Wi84}, \cite{Ha86}). The difference between the
numerical and theoretical predictions is around $10\%$. For $\rho _{0}=1.33\times 10^{15}gcm^{-3}$
the maximum mass of the star is about $1M_{\odot }$. 

Generally our formulae (\ref{12}) and (\ref{13})
underestimate the maximum values of the mass and radius because we have assumed
that the density inside the star is uniform. It is obvious that near the surface the
density is much lower than at the center of the compact object. Due to the approximations and simplifications
used to derive the basic expressions, reflected mainly in the uncertainties in the
exact value of the coefficient $R_{0}$,  Eqs. (\ref{12}) and (\ref{13}) cannot provide high
precison numerical values of the maximum mass and radius for linear EOS stars, which must be
obtained by numerically integrating the gravitational field equations.

For the  the linear EOS, $M_{max }$ and $R_{max}$ depend mainly on the
fundamental constants $c$ and $G$ and on the zero pressure density $\rho
_{0} $ (the bag constant). The Chandrasekhar expressions for the same
physical parameters involve two more fundamental constants, $\hbar $
and the mass of the electron or neutron.

For quark stars, usually one assume
they are composed of a three-flavour system of massless quarks, confined in
a large bag. Hence the mass of the quark cannot play any role in the mass
formula. But the linear EOS with arbitrary $a$ can describe quark matter
with non-zero quark masses (the mass of the strange quark $m_{s}\approx
200MeV$), forming a degenerate Fermi gas (\cite{Go00}, \cite{Zd00}). Therefore
this system should also be described by the same formulae as white dwarfs or
neutron stars, not only by Eqs. (\ref{12}) - (\ref{13}). Generally $\rho _{0}$ is
 a function of the mass of the strange quark, so this mass
implicitly appears in the expression of the maximum mass and radius. But on
the other hand we can assume that the Chandrasekhar limit also applies to
quark stars with the baryon mass substituted by an effective quark mass $%
m_{qeff}$, representing the minimum mass of the quark bubbles composing the
star. Hence we must have
\begin{equation}\label{15}
\left( \frac{\hbar c}{G}m_{qeff}^{-4/3}\right) ^{3/2}\sim \frac{c^{3}}{G}%
\frac{1}{\sqrt{\pi G\rho _{0}}}.  
\end{equation}

Eq. (\ref{15}) leads to the following expression of the effective mass of
the ''elementary'' quark bubble: 
\begin{equation}\label{16}
m_{qeff}\sim \left( \frac{\hbar \rho _{0}^{1/3}}{c}\right) ^{3/4}.
\end{equation}

The effective quark mass is determined only by elementary particle physics
constants and is independent of $G$. From its construction $m_{qeff}$ should
be relevant when the system is quantum mechanical and involves high
velocities and energies.
With respect to a scaling of the zero pressure density of the form $\rho
_{0}\rightarrow k\rho _{0}$, the effective quark mass has the scaling
behavior $m_{qeff}\rightarrow m_{qeff}k^{1/4}$, similar to the scaling of
the strange quark mass (\cite{Zd00}).

For $\rho _{0}=4\times 10^{14}gcm^{-3}$ we obtain $%
m_{qeff}\sim 3.63\times 10^{-25}g\approx 204MeV$. For $\rho _{0}=1.33\times 10^{15}gcm^{-3}$
Eq. (\ref{16}) gives $m_{qeff}\sim 4.9\times 10^{-25}g\approx 275MeV$.
The mass given by Eq. (\ref{16}) can be considered as the minimum
mass of the stable quark bubble. It is of the same order of magnitude as
the mass $m_{s}$ of the strange quark. Therefore the Chandrasekhar limit applies
also for quark stars if we take $m_{qeff}$ for the mass of the elementary constituent of the star
.

In the present paper we have considered the maximum mass and radius of strange stars
in the linear approximation of the equation of state and the dependence of these
quantities on the parameter $a$ has been found. We have also pointed out the
existence of scaling relations for the maximum radius of strange stars, an aspect that
has not been mentioned in previous investigations (\cite{Wi84}, \cite{Ha86}, \cite{Ba00}, \cite{Zd00}).
Our formulae also lead to the transformation relations for the maximum mass and
radius of strange stars with respect to separate and simultaneous scaling of the parameters
$a$ and $\rho _{0}$. On the other hand the possibility of estimation of the mass of the strange
quark from general astrophysical considerations can perhaps give a better understanding
of the deep connection between micro- and macro-physics.

{\it Acknowledgements.} This work is partially supported by a RGC grant of Hong Kong Government and T.H.
is supported by a studentship of the University of Hong Kong. The authors are very
grateful to the anonymous referee whose comments helped to improve an earlier version of the
manuscript.


\begin{thebibliography}{}

\bibitem[Alcock, Farhi \& Olinto 1986]{Al86}  Alcock C., Farhi E. \& Olinto A. 1986, Astrophys. J., 310, 261 

\bibitem[Bannerjee, Ghosh \& Raha 2000]{Ba00}  Bannerjee S., Ghosh S. K. \&  Raha S. 2000, J. Phys. G: Nucl. Part. Phys., 26, L1 

\bibitem[Cheng, Dai \& Lu 1998]{Ch98}  Cheng K. S., Dai Z. G. \&  Lu T. 1998, Int. J. Mod. Phys. D, 7, 139 

\bibitem[Cheng \& Harko 2000]{ChHa00} Cheng K. S. \&  Harko T. 2000, Phys. Rev. D, 62, 083001 

\bibitem[Dey et al. 1998]{De98} Dey  M., Bombacci  I., Dey J., Ray S. \&  Samanta  B. C. 1998, Phys. Lett. B, 438, 123  

\bibitem[Farhi \& Jaffe 1984]{Fa84} Farhi E. \&  Jaffe  R. L. 1984, Phys. Rev. D, 30, 2379 

\bibitem[Frieman \& Olinto 1989]{Fr89} Frieman J. A. \&  Olinto A. 1989, Nature, 341, 633 

\bibitem[Glendenning 1996]{Gl96} Glendenning N. K. 1996, Compact stars: nuclear physics,
particle physics and general relativity, Springer, New York 

\bibitem[Gondek-Rosinska et al. 2000] {Go00} Gondek-Rosinska D., Bulik T., Zdunik L., Gourgoulhon  E., Ray  S., Dey J. \& Dey M. 2000, Astron. Astrophys., 363, 1005

\bibitem[Haensel, Zdunik \& Schaeffer 1986]{Ha86} Haensel  P., Zdunik J. L. \& Schaeffer  R. 1986, Astron. Astrophys., 160, 121 

\bibitem[Haensel \& Zdunik 1989]{Ha89} Haensel  P. \&  Zdunik J. L. 1989, Nature, 340, 617 

\bibitem[Lattimer et al. 1990]{La90} Lattimer J. M., Prakash  M., Masaak D. \& Yahil A. 1990, Astrophys. J., 355, 241 

\bibitem[Prakash, Baron \& Prakash 1990] {Pr90} Prakash M., Baron  E. \& Prakash M. 1990, Phys. Lett. B, 243, 175 

\bibitem[Rhoades \& Ruffini 1974]{Rh74}  Rhoades C. E. \&  Ruffini R. 1974, Phys. Rev. Lett., 32, 324 

\bibitem[Shapiro \& Teukolsky 1983]{ShTe83} Shapiro S. L. \&  Teukolsky S. A. 1983, Black Holes, White Dwarfs, and Neutron Stars, John Wiley \& Sons, New York 

\bibitem[Witten 1984]{Wi84}  Witten  E. 1984, Phys. Rev. D, 30, 272 

\bibitem[Zdunik 2000]{Zd00} Zdunik J. L. 2000, Astron. Astrophys., 359, 311 

\end{thebibliography}
\end{document}